\documentclass[9pt]{article}
\usepackage{extsizes}

\usepackage{spconf,amsmath,graphicx}
\usepackage{amsmath}
\usepackage{amssymb}
\usepackage{arydshln}
\usepackage{url}
\usepackage{dsfont}

\let\oldbibliography\thebibliography
\renewcommand{\thebibliography}[1]{%
  \oldbibliography{#1}%
  \setlength{\itemsep}{0.0em}%
}

\usepackage{etoolbox}
\newtoggle{prepub}
\toggletrue{prepub}
\iftoggle{prepub}{
    \usepackage{fancyhdr, textcomp}
    \fancypagestyle{IEEE_Copyright_footer}{%
        \fancyhf{}
        
        \fancyfoot[L]{\fontsize{9}{11} \selectfont {\textcopyright} 2021 IEEE. Personal use of this material is permitted. Permission from IEEE must be obtained for all other uses, in any current or future media, including reprinting/republishing this material for advertising or promotional purposes, creating new collective works, for resale or redistribution to servers or lists, or reuse of any copyrighted component of this work in other works.}
    }
    \pagestyle{empty}
}
\title{Speech Emotion Recognition using Semantic Information}
\name{Panagiotis Tzirakis$^1$, Anh Nguyen$^1$, Stefanos Zafeiriou$^1$, Bj{\"o}rn W.\ Schuller$^{1,2}$}
\address{$^1$ GLAM -- Group on Language, Audio, \& Music, Imperial College London, UK\\
  $^2$ EIHW -- Chair of Embedded Intelligence for Health Care and Wellbeing, University of Augsburg, Germany \\ email: {panagiotis.tzirakis12@imperial.ac.uk} }

\begin{document}

\maketitle
\iftoggle{prepub}{\thispagestyle{IEEE_Copyright_footer}}{\thispagestyle{empty}}

\begin{abstract}
Speech emotion recognition is a crucial problem manifesting in a multitude of applications such as human computer interaction and education. Although several advancements have been made in the recent years, especially with the advent of Deep Neural Networks (DNN), most of the studies in the literature fail to consider the semantic information in the speech signal. In this paper, we propose a novel framework that can capture both the semantic and the paralinguistic information in the signal. In particular, our framework is comprised of a semantic feature extractor, that captures the semantic information, and a paralinguistic feature extractor, that captures the paralinguistic information. Both semantic and paraliguistic features are then combined to a unified representation using a novel attention mechanism. The unified feature vector is passed through a LSTM to capture the temporal dynamics in the signal, before the final prediction. To validate the effectiveness of our framework, we use the popular SEWA dataset of the AVEC challenge series and compare with the three winning papers. Our model provides state-of-the-art results in the valence and liking dimensions. \footnote{Code available here: \url{https://github.com/glam-imperial/semantic_speech_emotion_recognition}}
\end{abstract}

\begin{keywords}
emotion recognition, deep learning, semantic, paralinguistic, audiotextual information
\end{keywords}

\section{Introduction}

Automatic affect recognition is a vital component in human-to-human communication affecting our social interaction, perception among others~\cite{picard2000affective}. 
In order to accomplish a \textit{natural} interaction between human and machine, intelligent systems need to recognise the emotional state of individuals. However, the task is challenging, as human emotions lack of temporal boundaries and different individuals express emotions in different ways~\cite{anagnostopoulos2015features}. In addition, emotions are expressed through multiple modalities. Over the past two decades, a plethora of systems have been proposed that utilise several modalities such as physiological signals, facial expression, speech, and text~\cite{TZIRAKIS2019387, kollias2019deep, schuller2018speech, stappen2020muse, zhang2020emotion}. To achieve an accurate emotion recognition system, it is important to consider multiple modalities, as complementary information exists among them~\cite{TZIRAKIS2019387}.

Current studies exploit Deep Neural Networks (DNNs) to model affect using multiple modalities~\cite{tzirakis2017end, albanie2018emotion, tzirakis2018end2you}. 
Two modalities that have been extensively used for the emotion recognition task are speech and text~\cite{yoon2018multimodal, mantyla2018evolution}. Whereas the speech signal provides low-level characteristics of the emotions (e.\,g., prosody), text provides high-level (semantic) information (e.\,g., the words ``love'' and ``like'' carry strong emotional content). To this end, several systems have shown that integrating both modalities, strong performance gains can be obtained~\cite{yoon2018multimodal}. 

However, one may argue that the textual information is redundant, as it is already included in the speech signal, and as such semantic information can be captured using only the speech modality. To this end, we propose an audiotextual training framework, where the text modality is used during training, but discarded during evaluation. In particular, we train Word2Vec~\cite{mikolov2013distributed} and Speech2Vec~\cite{chung2018speech2vec} models, and align their two embedding spaces such that Speech2Vec features are as close as possible with the Word2Vec ones~\cite{chung2018unsupervised}. In addition to the semantic information, we capture low-level characteristics of the speech signal by training a convolution recurrent neural network. The semantic and paralinguistic features are combined to a unified representation and passed through a long short-term memory (LSTM) module that captures the temporal dynamics in the signal, before the final prediction. 

To test the effectiveness of our model, we utilise the Sentiment Analysis in the Wild (SEWA) dataset, which was used in the Audio/Visual Emotion Challenge (AVEC) since 2017~\cite{avec2017}. The dataset provides three continuous affect dimensions: arousal, valence, and likability. Although the arousal and valence dimensions are easily integrated in a single network during the training phase of the models, the likability dimension can cause convergence and generalisation difficulties~\cite{avec2017, Chen:AVEC2017}. To this end, we propose to use a novel `disentangled` attention mechanism to fuse the semantic and paralinguistic features such that the information required per affect dimension is \textit{disentangled}. Our approach provides training stability, and, at the same time, increases the generalisability of the network during evaluation. We compare our framework with the three best performing papers of the competition~\cite{Dang:AVEC2017, Huang:AVEC2017, Chen:AVEC2017} in terms of concordance correlation coefficient ($\rho_c$)~\cite{schuller2018interspeech, tzirakis2021}, and show that our method provides state-of-the-art results for the valence and likability dimensions.

In summary, the main contributions of the paper are the following: (a) propose to use the acoustic speech signal to capture semantic information that exists in the text modality, (b) show how to disentangle the information in the network per affect dimensions for stable training and generalisability during the evaluation phase, and (c) produce state-of-the-art results in the valence and likability dimensions using the SEWA dataset. 

\begin{figure*}[!h]
    \centering
    \includegraphics[width=13.5cm]{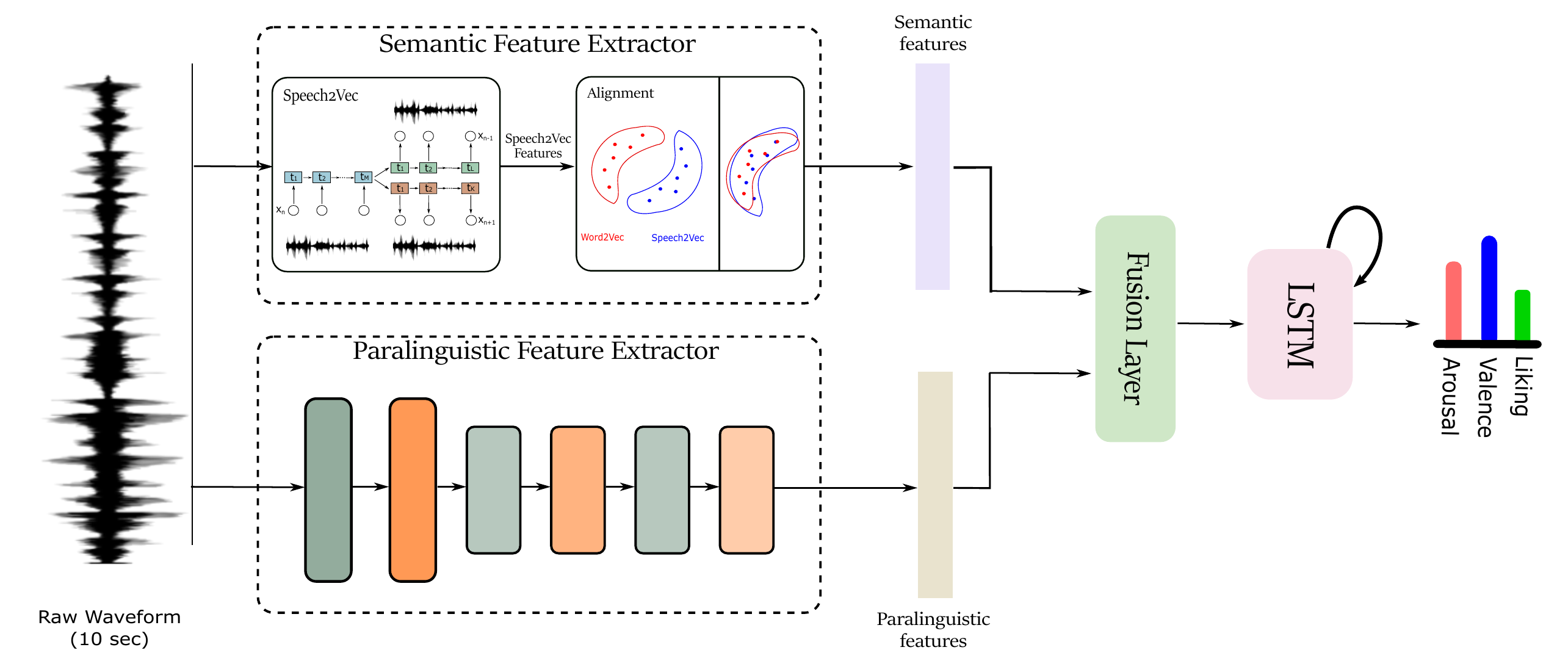}
    \caption{\textit{Our proposed model is comprised of two networks: (a) the semantic feature extractor, that extracts high-level features containing semantic information of the input, and (b) the paralinguistic feature extractor, that extracts low-level features containing paralinguistic information of the signal. Both feature vectors are passed through a fusion layer, that combines the information and extracts a unified representation of the input, that is then passed through a LSTM model for the final prediction.}}
    \label{proposed_method}
\end{figure*}

\section{Related work}

Several studies have been proposed in the literature for speech emotion recognition~\cite{trigeorgis2016adieu, tzirakis2018end, tarantino2019self}. For example, Trigeorgis et al.~\cite{trigeorgis2016adieu} utilised a convolution neural network to capture the spatial information in the signal, and a recurrent neural network for the temporal ones. In a similar study, Tzirakis et al.~\cite{tzirakis2018end} showed that utilising a deeper architecture with longer input window produces better results. In another study, Neumann et al.~\cite{neumann2017attentive} propose an attentive convolutional neural network (ACNN) that combines CNNs with attention. 

In the past ten years, a plethora of models have been proposed that incorporate more than one modality for the emotion recognition task~\cite{tzirakis2017end, han2019implicit, albanie2018emotion}. 
In particular, Tzirakis et al.~\cite{tzirakis2017end} uses both audio and visual information for continuous emotion recognition. Although this study produced good results, it utilises both modalities for the training and evaluation of the model. 
In a more recent study, Albanie et al.~\cite{albanie2018emotion} transfer the knowledge from the visual information (facial expressions) to the speech model. 
In another study, Han et al.~\cite{han2019implicit} proposed an implicit fusion strategy for audiovisual emotion recognition. In this study, both audio and visual modalities are used for the training of the model and only one for the evaluation of the model. 
% The difference with the proposed method is that we try to capture semantic information during training, and not distil knowledge.

\section{Proposed Method}

Our cross-modal framework can leverage the semantic (high-level) information (Sec.~\ref{semantic_features}) and the paralinguistic (low-level) dynamics in the speech signal (Sec.~\ref{par_features}). The low- and high-level feature sets are fused together using a novel attention fusion strategy (Sec.~\ref{fusion_strategies}) before feeding them to a one-layer LSTM module, to captures the temporal dynamics in the signal, for the ﬁnal frame-level prediction. Fig.~\ref{proposed_method} depicts the proposed method.

\subsection{Semantic Feature Extractor}
\label{semantic_features}

To capture the semantic information in the speech signal, we train Word2Vec and Speech2Vec models. The first model uses the text information to extract a semantic vector representation from a given word, whereas the second one the speech. We align their embedding spaces, similar to~\cite{chung2018unsupervised}, for semantically richer speech representations. Mathematically, we define the speech embedding matrix $S = [s_1, s_2, ..., s_m] \in \mathbb{R}^{m \times d_s}$ to be of $m$ vocabulary words with dimension $d_s$, and the text embedding matrix $T = [t_1, t_2, ..., t_n] \in \mathbb{R}^{n \times d_t}$ to be of $n$ vocabulary words with dimension $d_t$. Our goal is to learn a linear mapping $W \in \mathbb{R}^{d_t \times d_s}$ such that $WS$ is most similar to $T$. 
% To this end, we (i) learn an initial proxy of $W$ via domain-adversarial training, and (ii) using the best matching words, we build a synthetic dictionary mapped from $S_r$ to $T_r$ and use it to refine $W$. 

To this end, we learn an initial proxy of $W$ via domain-adversarial training. The adversarial training is a two-layer game where the generator tries, by computing $W$, to deceive the discriminator from correctly identifying the embedding space, and making $WS$ and $T$ as similar as possible. Mathematically, the discriminator tries to minimise the following objective:

\begin{equation}
    \begin{split}
        L_D(\theta_D|W) &= - \frac{1}{n}\sum_{i=1}^n{\textrm{log} P_{\theta_D}(speech = 1|Ws_i)} \\
                        &\quad - \frac{1}{m}\sum_{i=1}^m{\textrm{log} P_{\theta_D}(speech = 0|t_i)},
    \end{split}
\end{equation}

\noindent where $\theta_D$ are the parameters of the discriminator, and $P_{\theta_D}(speech=1|z)$ is the probability the vector $z$ originates from speech embedding.

On the other hand, the generator tries to minimise the following objective:

\begin{equation}
    \begin{split}
        L_G(W|\theta_D) &= - \frac{1}{n}\sum_{i=1}^n{\textrm{log} P_{\theta_D}(speech = 0|Ws_i)} \\ 
                        &\quad - \frac{1}{m}\sum_{i=1}^m{\textrm{log} P_{\theta_D}(speech = 1|t_i)}.
    \end{split}
\end{equation}

A limitation of the above formulation is that all embedding vectors are treated equally during training. However, words with higher frequency would have better embedding quality in the vector space than less frequent words. To this end, we use the frequent words to create a dictionary that specifies which speech embedding vectors correspond to which text embedding vectors, and refine $W$:

\begin{equation}
    \label{refine_eq}
    W^* = \underset{W}{\textrm{argmin}} \: ||WS_r - T_r||_F,
\end{equation}

\noindent where $S_r$ is a matrix built by \textit{k} speech vectors from $S$ and $T_r$ is a matrix built by \textit{k} vectors from $T$. The solution of Eq.~\ref{refine_eq} is obtained from the singular value decomposition of $S_rT_r^T$, i.\,.e., $SVD(S_rT_r^T) = U \Sigma V^T$.

\subsection{Paralinguistic Feature Extractor}
\label{par_features}

Our paralinguistic feature extraction network is comprised of three 1-D CNN layers with a rectified linear unit (ReLU) as activation function, and max-pooling operations in-between. Both convolution and pooling operations are performed on the time domain, using the raw waveform as input. Inspired by our previous work~\cite{tzirakis2018end}, we perform convolution with small kernel size and stride of one, and a large kernel and stride size for the max-pooling.
Table~\ref{tab:network} shows the architecture of the network. 

\begin{table}
\centering
\begin{tabular}{l c c c}  \hline
\textbf{Layer}       & \textbf{Kernel/Stride}   & \textbf{Channels} & \textbf{Activation} \\ \hline
Convolution & $8/1$ & $50$ & ReLU \\
Max-pooling & $10/10$ & --- & --- \\
Convolution & $6/1$ & $125$ & ReLU \\
Max-pooling & $5/5$ & --- & --- \\
Convolution & $6/1$ & $125$ & ReLU \\
Max-pooling & $5/5$ & --- & --- \\ \hline
\end{tabular}
\caption{Paralinguistic feature extractor. Shown are the layer type, kernel/stride size, channels size, and activation function.}
\label{tab:network}
\end{table}

\subsection{Fusion Strategies}
\label{fusion_strategies}

Our last step is to fuse the semantic ($\mathbf{x}_s \in \mathds{R}^{d_s}$) and paralinguistic ($\mathbf{x}_p \in \mathbb{R}^{d_p}$) speech features, before feeding them to the LSTM. This is performed with two strategies: (i) concatenation, (ii) `disentangled` attention mechanism.

\textit{Concatenation}. 
The first approach is a standard feature-level fusion, i.\,e., a simple concatenation of the feature vectors. Mathematically, $\mathbf{x}_{fusion} = [\mathbf {x}_s, \mathbf {x}_p]$. 

\textit{Disentangled attention mechanism}. 
For our second approach, we propose using attention mechanism to fuse the two modalities. To this end, we perform a linear projection for each of the feature sets such that they are in the same vector space (with dimension $d_u$): \\

\begin{equation}
\begin{split}
    \tilde{\mathbf{x}}_s &= \mathbf{W}_s\mathbf{x}_s + b_s, \\
    \tilde{\mathbf{x}}_p &= \mathbf{W}_p\mathbf{x}_p + b_p,
\end{split}
\end{equation}
where $\mathbf{W}_s\in\mathds{R}^{d_u\times d_s}$, $\mathbf{W}_p\in\mathds{R}^{d_u\times d_p}$ are projection matrices for the semantic and paralinguistic feature sets, respectively.

We fuse these features using attention mechanism, i.\,e.,

\begin{equation}
    \begin{split}
        Attention(\tilde{\mathbf{x}}_s, \tilde{\mathbf{x}}_t) &= \alpha_s \tilde{\mathbf{x}}_s + \alpha_p \tilde{\mathbf{x}}_p, \\
        {\alpha}_i &= softmax(\frac{\tilde{\mathbf{x}}_i \mathbf{q}_i}{\sqrt{d_u}}), \\
    \end{split}
\end{equation}
\noindent
where $\mathbf{q} \in \mathds{R}^{d_u}$ is learnable vector that attends to different features.

At this point, we use three fully-connected (FC) layers with linear activation of same dimensionality on top of the output obtained from first attention layer, i.\,e., 

\begin{equation}
    \begin{split}
        {\mathbf{a}} &= \mathbf{W}_a\tilde{\mathbf{x}}_{sp} + b_a \\
        {\mathbf{v}} &= \mathbf{W}_v\tilde{\mathbf{x}}_{sp} + b_v, \\
        {\mathbf{l}} &= \mathbf{W}_l\tilde{\mathbf{x}}_{sp} + b_l,
    \end{split}
\end{equation}
\noindent
where $\{\mathbf{W}_a, \mathbf{W}_l, \mathbf{W}_v\} \in\mathds{R}^{d_u\times d_u}$ are projection matrices.

We choose to use three FC layers such that the information flow per emotional dimension (i.\,e., arousal, valence, and liking) in the network is disentangled. The intuition here is that by adding three additional dense layers, we hope that each of these projections could learn features that suit best for a dimension in our emotion space. In case of a higher number of outputs, more FC layers can be used. 

To fuse the information of the `disentanngled' vector spaces, we apply an attention layer so that each suited feature set could attend to one another and produce an enriched fusion feature output for final prediction. In particular, we, first, apply attention on ${\mathbf{a}}$ and ${\mathbf{l}}$; and, finally, on the result with ${\mathbf{v}}$, i.\,e., 

\begin{equation}
    \begin{split}
        \mathbf{z} &= Attention({\mathbf{a}}, {\mathbf{l}}) \\
        {\mathbf{x}}_{fusion} &= Attention(\mathbf{z},{\mathbf{v}}). \\
    \end{split}
\end{equation}

%$Attention(\tilde{\mathbf{x}}_a, \tilde{\mathbf{x}}_t)=\alpha_s \tilde{\mathbf{x}}_s + \alpha_p \tilde{\mathbf{x}}_p$, with ${\alpha}_i = softmax(\frac{\sum_{k}{\tilde{\mathbf{x}}_k \mathbf{\alpha}_k}}{\sqrt{d_u}})$.

\section{Dataset}

We test the performance of our proposed framework on a time-continuous emotion recognition dataset for real-world environments. In particular, as outlined, we utilise the Sentiment Analysis in the Wild (SEWA) dataset that was used in the AVEC 2017 challenge~\cite{avec2017}. The dataset consists of `in-th-wild' audiovisual recordings that were captured from web-cameras and microphones from $32$ pairs (i.\,e., $64$ participants) that watched a $90$\,sec commercial visual and discussed it with their partner for maximum of $3$\,min. It provides three modalities, namely, audio, visual, and text, for three emotional dimensions: arousal, valence, and liking. The dataset is split into 3 partitions: training ($17$ pairs), development ($7$ pairs), and test ($8$ pairs), and was annotated by $6$ German-speaking annotators ($3$ female, $3$ male). 

\section{Experiments}

\subsection{Experimental Setup}

For training the models, we utilised the Adam optimisation method ~\cite{adam}, and a fixed learning rate of $10^{-4}$ throughout all experiments. We used a mini-batch of $25$ samples with sequence length of $300$, and a dropout~\cite{srivastava2014dropout} with $p=0.5$ for all layers except the recurrent ones to regularise our network. This step is important, as our models have a large amount of parameters and not regularising the network makes it prone on overfitting on the training data. In addition, the LSTM network we use in the training phase is trained with a dropout of $0.5$ and a gradient norm clipping of $5.0$. Finally, we segment the raw waveform into $10$\,sec long sequences with sampling rate of $22\,050$\,Hz. Hence, each sequence corresponds to a $22\,0500$-dimension vector.

\subsection{Objective Function}
\label{obj_fun}

Our objective function is based on the Concordance Correlation Coefficient ($\rho_c$) that was also used in the AVEC 2017 challenge. $\rho_c$ evaluates the agreement level between the predictions and the gold standard by scaling their correlation coefficient with their mean square difference. Mathematically, the the concordance loss ${J}_{c}$ can be defined as follows:

\begin{equation}
    \begin{split}
        \mathcal{L}_{c} = 1 - \rho_c = 1 - \dfrac{2 \sigma_{xy}^2}{\sigma_x^2 + \sigma_y^2 + (\mu_x - \mu_y)^2},
    \end{split}
\label{eq:ccc_loss}
\end{equation}

\noindent where $\mu_{x} = \mathbb{E}(\mathbf x)$, ${\mu_{y} = \mathbb{E}(\mathbf y)}$, ${\sigma_x^2 = \mbox{var}(\mathbf x)}$, $\sigma_y^2 = \mbox{var}(\mathbf y)$, and $\sigma^2_{xy} = \mbox{cov}(\mathbf x, \mathbf y)$.

Our end-to-end network is trained to predict the arousal, valence, and liking dimensions, and as such, we define the overall loss as follows, $\mathcal{L} = ({\mathcal{L}_{c}^{a} + \mathcal{L}_{c}^{v} + \mathcal{L}_{c}^{l})/3},$
% \begin{equation}
%     \mathcal{L} = \frac{\mathcal{L}_{c}^{a} + \mathcal{L}_{c}^{v} + \mathcal{L}_{c}^{l}}{3},
% \end{equation}
% \noindent
where $\mathcal{L}_{c}^{a}$, $\mathcal{L}_{c}^{v}$, and $\mathcal{L}_{c}^{v}$ are the concordance loss of the arousal, valence, and liking dimensions, respectively, contributing equally to the loss.

\subsection{Ablation Study}

\subsubsection{Comparing Vector Spaces}

We test the performance of both the semantic and paralinguistic networks, trained independently, and trained jointly, to show the beneficial properties of our proposed framework. Table~\ref{ablation_results} depicts the results in terms of $\rho_c$ on the development set of the SEWA dataset. We observe that Word2Vec produces slightly better results than Speech2Vec. However, after aligning their embedding spaces, the aligned Speech2Vec has higher performance than Word2Vec, indicating both that the refinement process makes speech embedding similar to the word ones, and that paralinguistic information exists in the model. Finally, the paralinguistic network, although it produces worse results than the aligned Speech2Vec model, provides the best results for the arousal dimension.

\begin{table}[h]
\begin{tabular}{l| r r r | r}
    \hline
    \hline
    \textbf{Model} & \textbf{Arousal} & \textbf{Valence} & \textbf{Liking} & \textbf{Avg}\\
    \hline
    Word2Vec & .434 & .513 & .208 & .385 \\ \hdashline
    Speech2Vec & .433 & .470 & .182 & .362 \\ 
    Align Speech2Vec & .453 & .452 & .257 & .387 \\ \hdashline 
    Paralinguistic  & .508 & .436 & .154 & .366 \\ 
    \hline
    \hline
\end{tabular}
\caption{\textit{SEWA dataset results (in terms of $\rho_c$) of the Word2Vec, Speech2Vec, aligned Speech2Vec and paralinguistic models, on the development set.}}
\label{ablation_results}
\end{table}

\subsubsection{Fusion Strategies}

We further explore the effectiveness of the attention fusion strategy compared to the simple concatenation. For our experiments we utilised both semantic and paralinguistic deep network models of the proposed method. Table~\ref{attention_ablation_results} depicts the results in terms of $\rho_c$ on the development set of the SEWA dataset. We observe that the attention method performs superior to the other one on all emotional dimensions, indicating the effectiveness of our approach to model the three emotional dimensions by projecting them to three different spaces before fusing them together with attention.

\begin{table}[!h]
\begin{tabular}{l| r r r | r}
    \hline
    \hline
    \textbf{Fusion strategy} & \textbf{Arousal} & \textbf{Valence} & \textbf{Liking} & \textbf{Avg}\\
    \hline
    Concatenation & .427 & .428 & .306 & .387 \\
    Disentangled attention & .499 & .497 & .311 &  \textbf{.435}\\ 
    \hline
    \hline
\end{tabular}
\label{attention_ablation_results}
\caption{\textit{SEWA dataset results (in terms of $\rho_c$) of the various fusion methods (i.\,e., concatenation, attention, and hierachical attention) in the development set.}}
\end{table}

\subsection{Results}

We compare our proposed framework with the winning papers of the AVEC 2017 challenge. As our model utilises only the audio modality during evaluation, we show, for fairness of comparison, the results of these studies using the audio information. Table~\ref{final_results} depicts the results. First, we observe that our approach provides the best results in the valence dimension with high margin, and the second best in the arousal one.
We should note, however, that the network from Huang et al.~\cite{Huang:AVEC2017} was pretrained on $300$ hours of a spontaneous English speech recognition corpus before fine-tuning it to the SEWA dataset. In addition to the features of the network, they also utilise several hand-engineered features. Second, we observe that our approach provides the highest performance in the likability dimension. Our method is able to generalise on this dimension compared to Chen et al~\cite{Chen:AVEC2017} whose performance drops significantly compared to its performance on the development set. Finally, we should note the high generalisation capability of our approach to model all three emotional dimensions, indicating the effectiveness of the proposed disentangled attention mechanism strategy.
% Similarly, Chen et al.~\cite{Chen:AVEC2017} also utilise several hand-crafted features. In contrast to these methods, our model uses the raw waveform for the final prediction. 
% Finally, we should point out the high generalisation capability of our approach to model the liking dimension, which outperforms the other methods, indicating the effectiveness of the proposed disentangled attention mechanism strategy.

\begin{table}
\begin{tabular}{l|r r r}
    \hline
    \hline
    \textbf{Method} &  \textbf{Arousal} & \textbf{Valence} & \textbf{Liking}  \\
    \hline
    Baseline~\cite{avec2017} & .225 (.344) & .224 (.351) & -.020 (.081) \\
    Dang et. al.~\cite{Dang:AVEC2017} & .344 (.494) & .346 (.507) & --- (---) \\ 
    Huang et. al.~\cite{Huang:AVEC2017} & \textbf{.583} (.584) & .487 (.585) & --- (---) \\ 
    Chen et. al. \cite{Chen:AVEC2017} & .422 (.524) & .405 (.504) & .054 (.273) \\ 
    % Chen et. al.~\cite{Chen:AVEC2017} & .437 (.527) & .494 (.447) & .024 (.354) \\ 
    \hline
    \hline
    % \hdashline
    Proposed & .429 (.499) & \textbf{.503} (.497) & \textbf{.312} (.311) \\
    \hline
    \hline
\end{tabular}
\caption{\textit{SEWA dataset test result (in terms of $\rho_c$) of our proposed fusion model compared with the winning models in AVEC 2017. In parenthesis are the performances obtained on the development set.  A dash is inserted if the results could not be obtained.}}
\label{final_results}
\end{table}

\section{Conclusions}

In this paper, we propose a training framework using audio and text information for speech emotion recognition. In particular, we use Word2Vec and Speech2Vec models, and align their embedding spaces for accurate semantic feature extraction using only the speech signal. We combine the semantic and paralinguistic features using a novel attention fusion strategy that first disentangles the information per emotional dimension, and then combines it using attention. The proposed model is evaluated on the SEWA dataset and produces state-of-the-art results on the valence and liking dimensions, when compared with the best performing papers submitted to the AVEC 2017 challenge.

In future work, we intend to use a single network to simultaneously capture the semantic and the paralinguistic information in the speech signal. This will result in simplifying, and at the same time, reducing the number of parameters of the model. Additionally, we intend to investigate the performance of the proposed method on categorical emotion recognition datasets.

\section{Acknowledgements}

%We would like to thank Yu-An Chung~\cite{chung2018unsupervised} for providing a pre-trained Speech2Vec model on the German language.
The support of the EPSRC Center for Doctoral Training in High Performance Embedded and Distributed Systems (HiPEDS, Grant Reference EP/ L016796/1) is gratefully acknowledged.

\newpage

\vfill\pagebreak

\bibliographystyle{IEEEbib}
\bibliography{mybib}

\end{document}